\documentclass[%
 aip,
 amsmath,amssymb,
 reprint,%
]{revtex4-1}

\usepackage{graphicx}
\usepackage{dcolumn}
\usepackage{bm}

\usepackage[utf8]{inputenc}
\usepackage[T1]{fontenc}
\usepackage{mathptmx}
\usepackage{etoolbox}
\usepackage{graphicx}
    \graphicspath{ {./25_nobar/} }
    \DeclareGraphicsExtensions{.png,.eps}
\usepackage{caption,subcaption}
\usepackage{dcolumn}
\usepackage{bm}

\usepackage{amsthm,amsmath,amsfonts,amssymb}
\usepackage{mathtools,mathrsfs}
\usepackage{setspace}
\usepackage{xcolor}
\usepackage{thmtools} 
\usepackage{diagbox,multirow} 

\usepackage[pdfdisplaydoctitle,colorlinks,breaklinks,urlcolor=blue,linkcolor=blue,citecolor=blue]{hyperref} 
\usepackage[nameinlink, capitalise]{cleveref}

\usepackage{todonotes}

\makeatletter
\def\@email#1#2{%
 \endgroup
 \patchcmd{\titleblock@produce}
  {\frontmatter@RRAPformat}
  {\frontmatter@RRAPformat{\produce@RRAP{*#1\href{mailto:#2}{#2}}}\frontmatter@RRAPformat}
  {}{}
}%
\makeatother
\begin{document}

\preprint{AIP/123-QED}

\title[A Statistical Learning Approach to Mediterranean Cyclones]{A Statistical Learning Approach to Mediterranean Cyclones}
\author{L. Roveri}
\affiliation{ 
Dipartimento di Matematica, Universit\`a di Pisa, Largo Pontecorvo 5, 56127 Pisa, Italia
}%
\author{L. Fery}
\affiliation{ 
Laboratoire
des Sciences du Climat et de l’Environnement, UMR 8212 CEA-CNRS-UVSQ,
Université Paris-Saclay, IPSL, 91191 Gif-sur-Yvette, France
}%
\affiliation{
SPEC, CEA, CNRS, Université Paris-Saclay, CEA Saclay, 91191 Gif-sur-Yvette, France
}
\author{L. Cavicchia}
\affiliation{ 
CMCC Foundation - Euro-Mediterranean Center on Climate
Change, Italy.
}%
\author{F. Grotto}
\affiliation{ 
Dipartimento di Matematica, Universit\`a di Pisa, Largo Pontecorvo 5, 56127 Pisa, Italia
}%
\email{leonardo.roveri at phd.unipi.it, lucas.fery at lsce.ipsl.fr, leone.cavicchia at cmcc.it, francesco.grotto at unipi.it}

\date{\today}

\begin{abstract}
Mediterranean cyclones are extreme meteorological events of which much less is known compared to their tropical, oceanic counterparts. 
The raising interest in such phenomena is due to their impact on a region increasingly more affected by climate change, but a precise characterization remains a non trivial task.
In this work we showcase how a Bayesian algorithm (Latent Dirichlet Allocation) can classify Mediterranean cyclones relying on wind velocity data, leading to a drastic dimensional reduction that allows the use of supervised statistical learning techniques for detecting and tracking new cyclones.
\end{abstract}

\maketitle

\begin{quotation}
Extreme metereological events mark changes of climate, and their description must rely on complex, non-autonomous systems.
This is especially true for the Mediterranean basin, a relatively small region characterized by complex topography and interactions with larger systems due to its transitional location. 
These peculiar features lead to a large variability of intensity and characteristics of extreme events such as Mediterranean cyclones, to which the present paper is devoted. 
Mediterranean cyclones tend to draw less attention in comparison with their oceanic counterparts. 
Still, despite their usually smaller size and intensity, and shorter lifetime, they can have severe socio-economic consequences on the region. 
A precise characterization of these phenomena, and consequently their detection and forecasting, is a non-trivial task: coherent structures are usually easily identifiable, but procedural definitions are often not satisfying and struggle in correctly identifying the phenomenon of interest. 

In this paper we propose a statistical learning approach for identification and classification of Mediterranean cyclones.
We analyze wind and pressure data by means of a Latent Dirichlet Allocation algorithm, identifying relevant patterns and leading to a drastic dimensional reduction, which then allows to use classical learning algorithms in order to efficiently recognize Mediterranean cyclones by training over a database of past events.
\end{quotation}

\section{Introduction}\label{sec:intro}

Weather in the Mediterranean basin constitutes a complex system whose modelling must involve challenging features such as: the fluid dynamics of the sea and atmosphere, the effects of a complex topography, the influence of larger neighbouring systems (e.g. the North Atlantic region). 
As a whole, the geophysical system is therefore a highly nonlinear, non-autonomous system subject to boundary effects.

Cyclogenesis is a distinguished feature of Mediterranean climate: Mediterranean cyclones are extreme meteorological events whose study has a rather long history, but for which a precise observational record -- in fact, even a precise operational identification procedure -- is still lacking. 
Meanwhile, the concurrence of regional climate change and the increase in intensity of Mediterranean cyclones\cite{reale2022future} and other extreme events such as heat waves\cite{darmaraki2019future} has made the detection and forecasting of these phenomena a compelling issue, raising the interest for the topic \cite{hatzaki2023medcyclones}.

Direct modelling of complex systems in a reliable way is a difficult task, especially if it is performed in order to describe tail events such as extreme climate phenomena.
On the other hand, climate sciences can profit from the rapid increase of available data and computational power of the last decades, which has opened the way for the application of data-driven methods and machine learning\cite{materia2023}. 
Here we are concerned with the latter: we will apply a Bayesian algorithm in order to extract information on the structure of Mediterranean cyclones without relying on an assigned theoretical model, that is we treat -- as a first step -- the detection of cyclones as an unsupervised learning problem.

Frihat et al.\cite{Frihat2021} first adapted a Latent Dirichlet Allocation (LDA) algorithm for classifying and generating snapshots of turbulent flows, with the objective of categorizing coherent structures generated in the fluid dynamic motion. 
As we will detail below, LDA is a Bayesian network devised to identify topics in text libraries with the goal of classifying single documents: the application to fluid dynamics is based on a discrete collection of snapshots of the fluid, interpreted as single texts composing the ``library'' of the whole, continuous fluid motion. 
Coherent vortex structures turn out to be identified as latent motifs in this framework.
Fery et al.\cite{Fery2022} applied the procedure in a geophysical context: instead of snapshots of the velocity field in a fluid dynamics experiment, pressure data over North Atlantic Ocean were analyzed in order to identify patterns in cyclonic/anticyclonic phenomena that lead to heat waves and cold spells. 

In considering Mediterranean cyclones, we applied LDA to discretized weather maps including pressure and wind data: grid points were considered as words in a text (the snapshot), and the associated (discretized) metereological data as the number of occurrences of that word.
Motifs identified in this way, combined together, were interpreted as generating meteorological configurations. 
Meteorological data was retrieved from the European Centre for Medium-Range Weather Forecasts (ECMWF) reanalysis, ERA5. 
A subsequent application of supervised deep learning methods allowed us to recognize cyclones by training on a database of previously identified Mediterranean cyclones\cite{Flaounas2023}. 
This allowed for precise and robust identification of the cyclones' presence and location.

The method we present foregoes an important aspect of cyclonic phenomena, that is their evolution in time. However, it has many remarkable advantages: implementation is easy, results are robust, and the overall computational cost is moderate. 
Involving the dynamics of cyclones in the statistical learning procedure might be crucial for forecasting, but it appears to be secondary in the scope of mere detection. Moreover, the technique we showcase can be regarded as a data analysis tool to be employed in combination with a numerical model for weather forecasting, i. e. it can be applied to weather maps produced by a numerical simulation.

We will first provide some background on geophysical and fluid dynamical aspects of the problem in \cref{sec:physics};
in \cref{sec:dic} we will describe the unsupervised learning algorithm (LDA) on which we rely for our study, and finally in \cref{sec:results} we will report results of supervised learning techniques applied to cyclone detection.

\section{Detecting Mediterranean Cyclones}\label{sec:physics}

Loosely speaking, a cyclone can be characterized as an ``atmospheric circulation persisting in a region over a period of time''. 
While suggestive, this is of course a completely unsatisfactory definition from a practical point of view: we now provide a summary description of the phenomena under consideration, 
and refer to the recent survey of Flaounas et al.\cite{Flaounas2022} for a comprehensive overview.

\subsection{Describing Mediterranean Cyclones}

Mediterranean cyclones are best introduced with a phenomenological description of their formation and evolution. 

In a fashion closely analogous to the formation of coherent structures observed in turbulent flows with typical length of many orders of magnitude smaller, \textbf{Mediterranean cyclogenesis} is typically due to the large scale forcing of polar jets triggering baroclinic instability\cite{raveh2017}. 
More specifically, potential vorticity filaments (somewhat analogous to 2D vortex thinning effects\cite{chen2006}) causing breaking of Rossby waves are typical precursors of Mediterranean cyclones. From a mathematical viewpoint, this can be described in terms of solutions of (linearized) quasi-geostrophic equations for potential vorticity $q$,
\begin{equation} \label{eq:qg}
\begin{aligned}
        & \begin{cases}
            \frac{\partial q}{\partial t}+u \cdot \nabla q=0 , \\
            q=\nabla^2 \psi+\beta y+\frac{\partial}{\partial z}\left(\frac{f_0^2}{N^2} \frac{\partial \psi}{\partial z}\right) 
        \end{cases} 
        \text{ for } 0<z<H, \\
        & \begin{cases}
            \frac{\partial q}{\partial t}+u \cdot \nabla q=0 , \\ 
            q = f_0 \frac{\partial \psi}{\partial z} 
        \end{cases} 
        \qquad \qquad \quad \ \ \text{ for } z=0, H,
\end{aligned}
\end{equation}
with $u=-\partial \psi / \partial y$, $f_0$ and $N$ respectively being the Coriolis and buoyancy parameter, and vertical velocity $w=0$. 
This PDE system possesses a purely zonal solution $u=U(y, z)$, with Rossby waves consisting in small perturbations $u=U+u'$. 
Additional time-dependent forcing on the right-hand side of the evolutionary parts of \eqref{eq:qg}, describing external effects, make the system non-autonomous and triggers instability of the zonal solution, that is small perturbations do not lead to a (quasi-)periodic solutions, but instead to the formation of regions in which potential vorticity concentrates. 
(We refer the reader to Vallis\cite{Vallis2017}, Chapter 9, for a thorough discussion). 

Interaction of the regional atmospheric system with a complex topography and larger scales of the atmospheric motion distinguishes Mediterranean cyclones from other extratropical cyclones.
Indeed, during the \textbf{cyclone life cycle} other relevant diabatic processes intervene (such as heat transfer at upper or lower atmosphere) and combine with baroclinic ones, therefore requiring an even more non-autonomous mathematical description.
The relation between baroclinic (convective) and diabatic forcing remains unclear: we thus refer to Flaounas et al.\cite{Flaounas2022} for an analysis of different components of potential vorticity evolution. 
As it was also mentioned, the effect of pronounced topographical differences also affects the evolution (although this cannot be regarded as a non-autonomous characteristic of the system), and it has been proposed that they play a relevant role also in cyclogenesis\cite{buzzi2020}.

With this insight, it is possible to prescribe quantitative characteristics and features in order to obtain a procedural definition of Mediterranean cyclones, which can then be applied for detection at different stages of their evolution. 
Variability inherent to subregions and seasons, typical spatial (horizontal and vertical) and velocity scales, different phases of the life cycle, the development of secondary, smaller-scale vortices and other features have all be taken into account in such procedural descriptions \cite{alpert1990climatological,Campins2011}.
Moreover, numerical simulations can effectively represent the most important features of cyclones, even in the case of extreme phenomena\cite{cavicchia2012,Fosser2024}.
Nevertheless, predictability of Mediterranean cyclones, and even detectability in the case of smaller ones, presents wide differences (for instance depending on the particular subregion under consideration)\cite{doiteau2024}. 

Let us also mention that the relation between Mediterranean cyclogenesis and climate change remains a mostly open problem. 
A diminished frequency of occurrence of cyclones has been reported\cite{Nissen2014}, but it is rather the effect on their intensity which calls for further investigation\cite{cavicchia2014,romero2013medicane}.

\subsection{Detecting Coherent Structures in Fluid Dynamics}

Alongside fluid dynamical instabilities, boundary layer effects are a fundamental mechanism in the onset of turbulence. 
It is in the context of wall turbulence that Kline, Reynolds et al.\cite{kline1967} first observed how coherent structures (\emph{vortices} or \emph{eddies}) originate in ``bursting events'' at the detachment of boundary layers. 

Frihat et al.\cite{Frihat2021} applied an LDA algorithm for detection (and generation) of coherent structures in snapshots of these turbulent flows. 
Starting from a dataset of single-time snapshots of the velocity field $u$ in a numerical simulation of wall turbulence in a channel, they considered a scalar field depending on velocity components that encodes a condition on the stress tensor relevant in the generation of coherent structures. 
Denoting by $f_1,\dots, f_n$ the single snapshots of the scalar field under consideration on the grid of points, which we denote by $x_1,\dots, x_N$, the authors discretized and normalized the values of the $f_i$'s so that the latter take values in a small set of natural numbers $\{0,\dots, k\}$. 
This allowed to apply ideas from latent semantic analysis:
indeed, the same framework can describe the number of occurrences $f_i(x_j)$ of the word $x_j$ inside the $i$-th document of a text corpus. The underlying idea is that each snapshot $f_i$ should be generated as a mixture of \emph{topics} or \emph{motifs} $z_1,\dots,z_m$ (for which a precise mathematical model is to be chosen, see the next Section) representing the various typical features that can be found in the corpus of snapshots (texts).

This approach is implemented as follows:
\begin{itemize}
    \item mathematical models (more specifically, parametric families of probability distributions) are chosen for defining the topics $z_1,\dots,z_m$ and their conditional relations with snapshots $f_i$ and grid points $x_j$;
    \item a Bayesian algorithm is used for determining the distributions of the topics $z_1,\dots,z_m$ and the probability distributions specifying how relevant each topic is in a given snapshot;
    \item since the latter step \emph{does not provide an interpretation} for the identified topics, this must rely on a different approach based on further information coming from previous knowledge of the physics of the system.
\end{itemize}

In the specific (numerical) experiment considered by Frihat et al.\cite{Frihat2021}, topics corresponded to coherent structures appearing at different locations and with different shapes in a 2D fluid, specifically in configurations of velocity fields in turbulent channel flow at a moderate Reynolds number.
While the mathematical model based on Dirichlet distributions is the same that we will use, the difference resides in the interpretation of the results, and the key contribution of the latent semantic analysis framework is a drastic dimensional reduction of the problem, which now involves few distinct objects ---the topics--- while irrelevant details of the starting vector fields have been statistically pruned out.

As opposed to wall turbulence, atmospheric cyclones constitute a much different phenomenon, if anything because their onset is mostly due to baroclinic instability. 
Nevertheless, an analogy can be drawn at least to the extent that the statistical learning approach proposed in the former case is well motivated also in the latter. 
Both phenomena involve fluids at low or moderate Reynolds number (almost inviscid fluids in wall turbulence, negligible fluid viscosity in atmospheric dynamics), and in both cases the physical description is at best phenomenological. 
In fact, concerning wall turbulence, we recall that Navier-Stokes equations are derived from a kinetic description by essentially relying on a linear response approximation\cite{ruelle1989},
and the derivation of boundary conditions from first principles is a complex problem.

\section{Looking for Mediterranean Cyclones in a Weather Dictionary}\label{sec:dic}

We now describe our application of LDA to the specific task of recognizing patterns in weather maps of the Mediterranean region, starting from which cyclones can be detected. We first recap the setup of the Bayesian learning algorithm we have employed.

\subsection{Latent Dirichlet Allocation}

Recall that we are assuming that a single snapshot $f_i$ is a non-negative scalar field over the grid $x_1,\dots,x_N$ taking integer values $\{0,\dots, k\}$. In applications differing from text analysis, one easily reduces to this setting by rescaling and discretizing the field:
it is safe to assume that the values of the field are bounded as one can safely neglect sporadic extreme values exceeding a high enough threshold. 
If the field under consideration present both positive and negative values, one can either analyze separately the positive and negative part, or simply rescale the whole interval of possible values to $[0,k]$ and then discretize.

The value $f_i(x_j)$ counts the number of times that the point $x_j$ has been ``activated'' in the snapshot, and the probabilistic model that we assume to underly the values $f_i(x_j)$ is the following. A \textbf{topic} (or \textbf{motif}) is a multinomial distribution over the grid points $x_i$ that models ``how likely is that $x_i$ is activated'' under such topic. 
The \textbf{Dirichlet distribution} is a probability distribution on the space of multinomial distributions that make Bayesian learning for this model tractable.

We recall that the Dirichlet distribution of order $M\geq 2$ and parameters $\gamma=(\gamma_1,\dots, \gamma_M)\in (0,\infty)^M$ has density 
\begin{equation*}
    p\left(x_1,\dots, x_M ; \gamma_1, \ldots, \gamma_M\right)=\frac{1}{B(\gamma)} \prod_{n=1}^M x_j^{\gamma_j-1},
\end{equation*}
over the $(M-1)$-dimensional simplex given by $y_1,\dots, y_M\in [0,1]$ with $\sum^M_{j=1} y_j=1$,
where the multivatiate beta function $B$ is given by
\begin{equation*}
    B(\gamma)=\frac{\prod_{j=1}^M \Gamma\left(\gamma_j\right)}{\Gamma\left(\sum_{j=1}^M \gamma_j\right)} .
\end{equation*}
The support of the distribution is to be identified with the set of multinomial distributions with parameters $y_1,\dots, y_M$. In our discussion, the \textbf{Bayesian prior} will always be the uninformative one, that is $\gamma_1=\dots=\gamma_M=1$.

A single snapshot $f$ then corresponds to a superposition of topics as follows:
\begin{itemize}
    \item a \emph{snapshot-topic} distribution is sampled from a Dirichlet distribution of order $m$ (the number of topics) and parameters $\alpha_1,\dots,\alpha_m$, we denote the associated probabilities with $p(z_h \mid f)$;
    \item for each topic $z_1,\dots, z_m$, a \emph{topic-site} distribution is sampled from a Dirichlet distribution of order $N$ (the number of sites) and parameters $\beta_1,\dots,\beta_N$, we denote the associated probabilities with $p(x_j\mid z_h)$;
    \item the ``total intensity'' $K=\sum^N_{i=1} f(x_i)$ is distributed over sites (that is, each $f(x_i)$ is determined) as follows: initializing $f\equiv 0$, for each $1,\dots K$, independently,
    \begin{itemize}
        \item a topic $z_h$ is sampled from the multinomial distribution $p(z_1 \mid f),\dots, p(z_m \mid f)$,
        \item a site $x_j$ is sampled from the multinomial distribution $p(x_1\mid z_h),\dots, p(x_N\mid z_h)$ and the value $f(x_j)$ incremented by one.
    \end{itemize}
\end{itemize}

For the sake of clarity, let us emphasize that the parameters specified a priori are:
\begin{itemize}
    \item the number of sites $N$,
    \item the ``total intensity'' $K$,
    \item the number $m$ of topics,
\end{itemize}
while the parameters that are the object of the Bayesian learning problem are the vectors $\alpha=(\alpha_1,\dots,\alpha_m)$ and $\beta=(\beta_1,\dots,\beta_N)$.

Given a new snapshot $f$, in order to compute the posterior distribution one needs to evaluate its likelihood given the prior with parameters $\alpha,\beta$, and in order to do so the probabilities $p(z_h\mid f)$ and $p(x_i\mid z_h)$ must be estimated. The task was tackled in Blei et al.\cite{blei2003latent} by means of a variational approach (minimization of Kullback-Leibler divergence) and the solution implemented in \texttt{gensim}\cite{rehurek_lrec} Python library, on which we have relied in the present work.

\subsection{Application of LDA to ERA5 wind data}

We selected the area spanning from 26° to 50° in latitude and from -10° to 45° in longitude, so to completely cover the whole Mediterranean region, and used the highest available data resolution, that is we chose data points collected hourly in the timeframe Jan. 1979 to Nov. 2020, on a grid of side length 0.25°.
We considered two variables which are expected to be the most relevant for characterizing a cyclone, namely sea-level pressure (slp) and wind intensity at 100m above mean sea level.

The procedure was first applied only to slp data, then only to wind intensity data and finally to their combination. Since LDA can be applied to a single (discretized) scalar field, the combined analysis was carried out in the supervised step consisting in locating cyclones from a given mixture of topics, to be described in the forthcoming Section.

Several trials were conducted on a reduced portion of the dataset. Topics in the analysis of slp data on its own were not sufficient to efficiently detect positions of cyclones: this is most probably due to the fact that many Mediterranean cyclones correspond to a pressure perturbation which is relevant at a local level, but negligible if compared with the cyclone-anticyclone trend over the whole region. This also resulted in redundant topics in the LDA output. Moreover, the combination of pressure data with wind data did not improve the efficacy of the procedure, so we eventually opted to use only wind intensity due to its better results. 

We relied on the implementation of the LDA algorithm of the \texttt{gensim}\cite{rehurek_lrec} library which was originally designed for text classification, and therefore accepts as training data arrays of small natural numbers corresponding to the number of occurrencies of words.
Considering a single weather snapshot as a document in a library, and the value of the scalar field under consideration at a point as the ``number of occurrencies'' of that point (regarded as a ``word''), we needed to discretize the ERA5 weather maps. 
Therefore, we preliminarily processed the data by considering the absolute intensity (euclidean norm) of the wind velocity (data were provided in the form of south-north and west-east wind velocities, in $m\cdot s^{-1}$). 
We then discarded all measurements below a certain threshold ($0.1$ $m\cdot s^{-1}$) and, 
after normalizing the maximum value on the grid (over the whole dataset) to be $20$,
we discretized the data to integer values. 
Thresholds were chosen with further empirical testing, and they are compatible with the experience in text document analysis for which the best results are obtained when the number of occurrences per word is relatively small.

Preprocessing led to develop a dataset of the form of a  $368'184\times21'437$ matrix of integers ranging between $0$ and $20$, where each row represented a document, i.e. measurements taken at a specific time, and each column stood for a word, i.e. a specific position on our grid. Indeed, the grid was reshaped as an array, and every (discretized and normalized) single measurement was intended as occurrences of a specific word.

After preprocessing, we split the dataset into multiple subsets, each spanning over $3$ years of measurements; this allowed us to run the algorithm on smaller datasets, keeping a better control over the whole procedure.
At the same time, due to the possibility of retraining the same LDA model with new data beacuse of the Bayesian setup, this did not led to a loss of precision.
We chose to iterate the algorithm for a maximum of $2000$ times (or until convergence) over chunks of $2^{13}$ rows, repeating the procedure $5$ times, always with the goal of maintaining a good balance between precision and computational costs.

\subsection{Results}

In general, there is no unambiguous methods to determine in advance the number of topics. This is clear if we think of the analogy with a large text corpus, for which determining exactly how many topics it covers is largely subjective. Reasonable guesses can only be made based on previous knowledge of the dataset.

In our context there was no prior knowledge allowing us to make such guess,
so we tried different choices ranging from $20$ to $30$ topics.
This is in slight contrast with the application of Fery et al.\cite{Fery2022}, in which pressure maps over the Atlantic were analyzed and knowledge on the typical pressure configurations over the region could be used to choose a number of topics and then to recognize the results.
How many (and what) typical structures should contribute in composing a Mediterranean cyclone is on the other hand a rather open question, even in its very meaning.

\begin{figure*}
	\includegraphics[width=4.24cm]{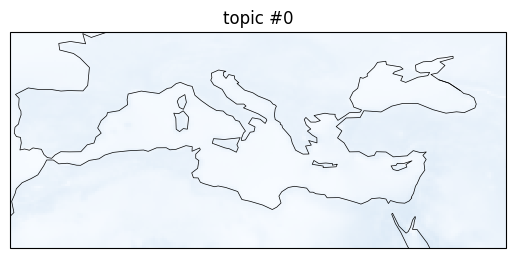}
	\includegraphics[width=4.24cm]{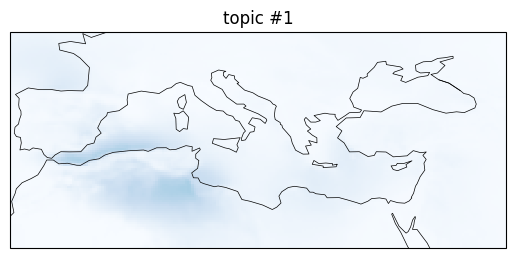}
	\includegraphics[width=4.24cm]{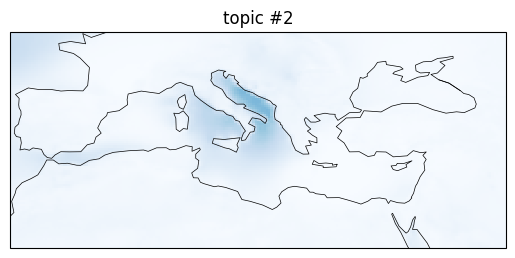}
	\includegraphics[width=4.24cm]{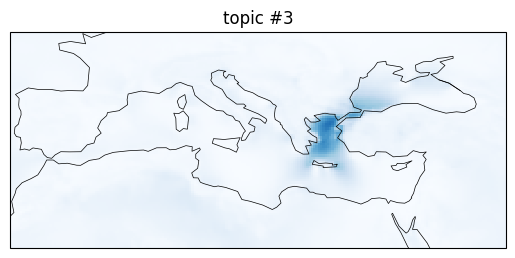}
	\includegraphics[width=4.24cm]{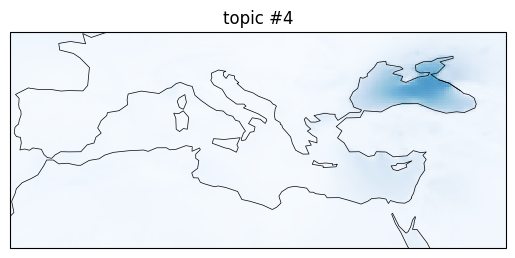}
	\includegraphics[width=4.24cm]{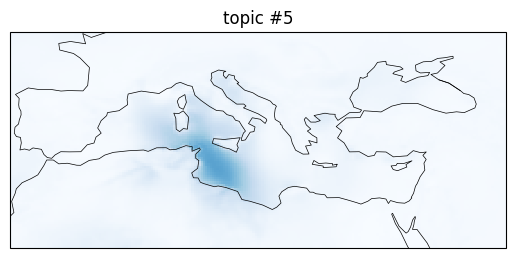}
	\includegraphics[width=4.24cm]{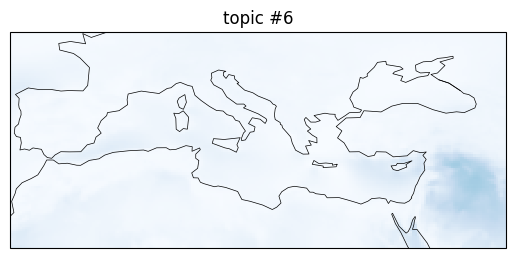}
	\includegraphics[width=4.24cm]{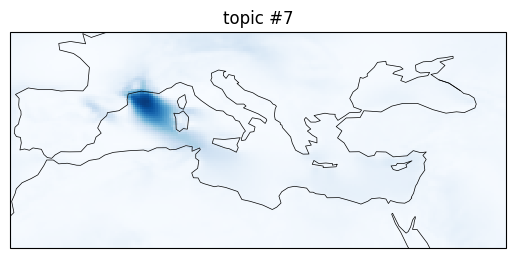}
	\includegraphics[width=4.24cm]{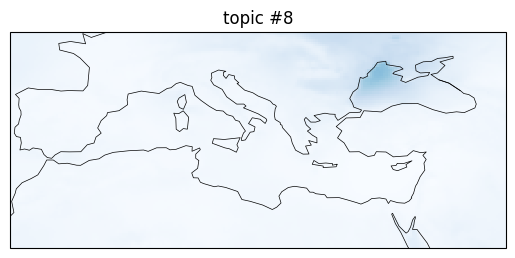}
	\includegraphics[width=4.24cm]{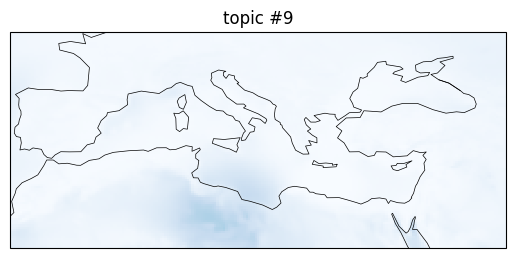}
	\includegraphics[width=4.24cm]{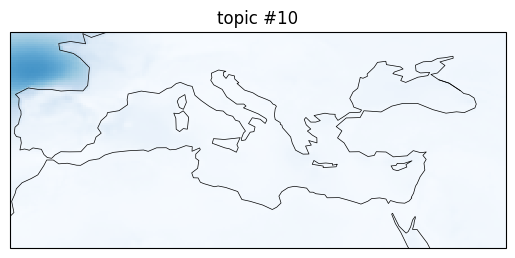}
	\includegraphics[width=4.24cm]{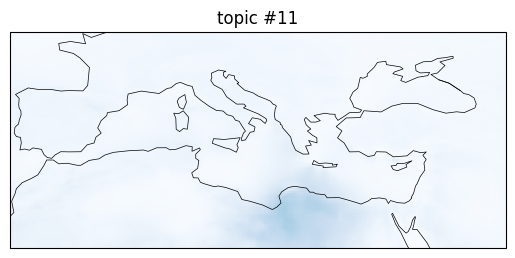}
	\includegraphics[width=4.24cm]{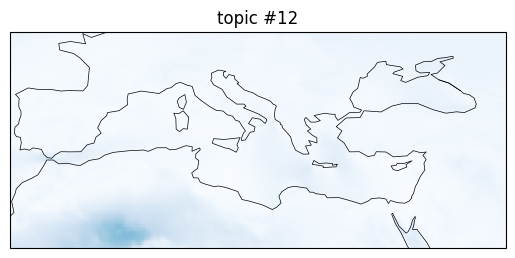}
	\includegraphics[width=4.24cm]{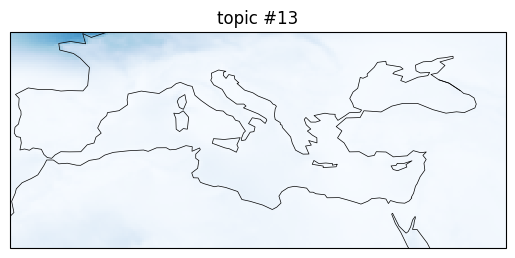}
	\includegraphics[width=4.24cm]{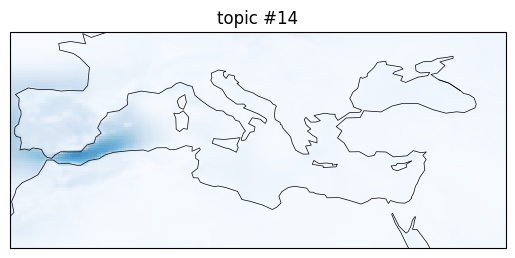}
	\includegraphics[width=4.24cm]{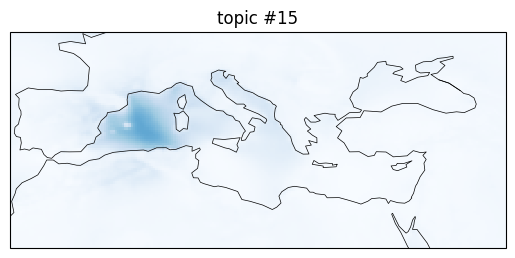}
	\includegraphics[width=4.24cm]{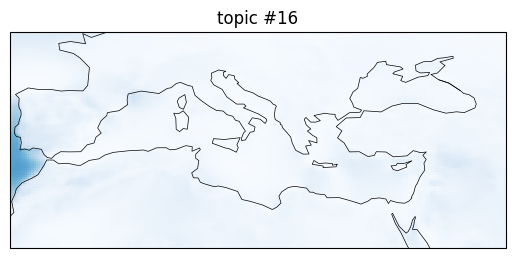}
	\includegraphics[width=4.24cm]{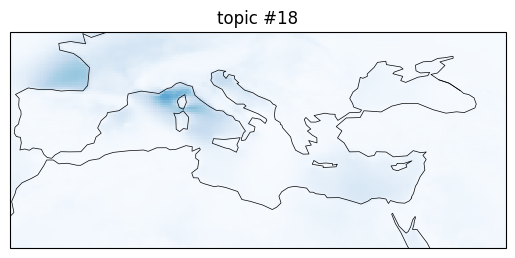}
	\includegraphics[width=4.24cm]{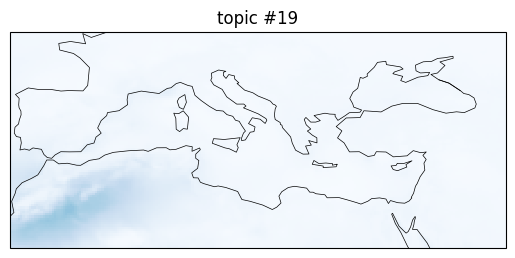}
	\includegraphics[width=4.24cm]{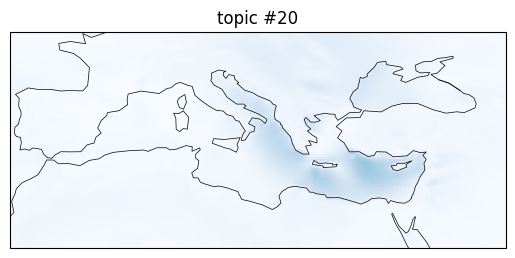}
	\includegraphics[width=4.24cm]{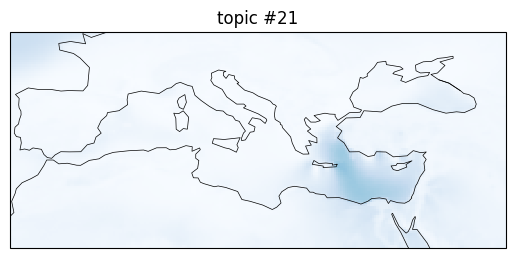}
	\includegraphics[width=4.24cm]{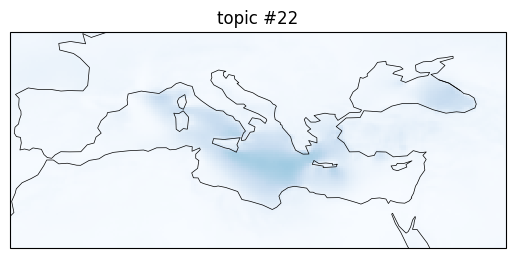}
	\includegraphics[width=4.24cm]{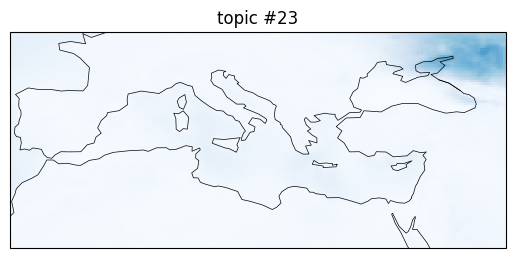}
	\includegraphics[width=4.24cm]{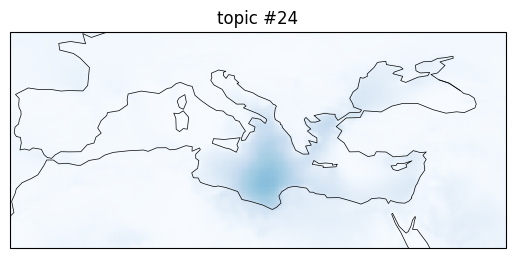}
    {\centering\includegraphics[width=8cm]{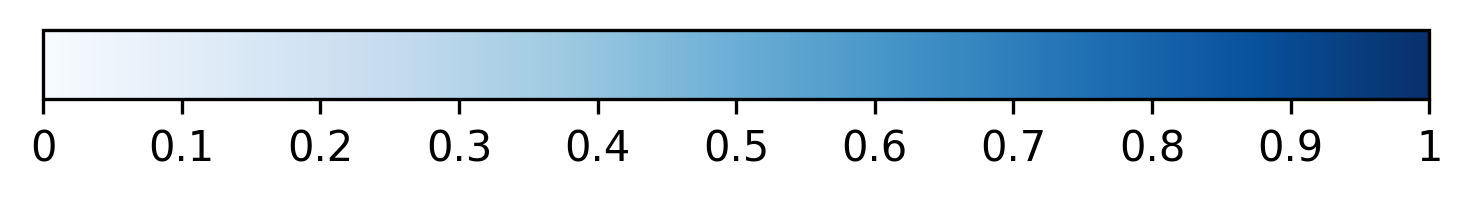}}
	\caption{\label{fig:motifs} The 25 topics identified by LDA as the ones generating Mediterranean weather configurations. Subfigures show the results of our algorithm. The number of motifs is chosen manually based on two principles: optimality with respect to statistical inference techniques for cyclones location given a weather (wind) map, and the fraction of area covered by the topics. Topic 17 is omitted since it is very similar to the first one.}
\end{figure*}

We reserved to make a definitive choice after comparing topic superpositions and the results of supervised learning techniques with topics as input data, with the idea that the more appropriate model should be the one yielding the most precise results for cyclone detection. 
We present in \cref{fig:motifs} the result of the LDA applied to our data, limited to the model that turned out to yield the better results. 

As can be seen in \cref{fig:motifs}, each motif tends to focus on a single geographical zone, with the only exceptions of n.$1$,$17$ and $18$. Of the latter, the first two present a very disperse scenario which was always obtained in at least a couple of topics no matter the assigned number of topics: that can be interpreted as a background profile collecting fluctuations at small scales.
Topic $17$ was therefore omitted in \cref{fig:motifs}, since it is essentially identical to the first one.
Topic $18$ spreads on two different zones, this however has not been an issue in our subsequent applications. 
Finally, let us observe that even if most topics are concentrated in a single localized subregion, different topics can be concentrated in the same area: this cautions against interpreting topics as ``generating cyclones in a certain zone''.

\section{(Supervised) Statistical Learning for Mediterranean Cyclones}\label{sec:results}

On its own, LDA does not provide an interpretation of the topics it identifies. 
In order to use the identified topics and detect the presence of cyclones we resorted to supervised learning algorithms using LDA topics as features and trained on the cyclone dataset of Flaounas et al.\cite{Flaounas2023}, collecting hourly latitude/longitude positions of Mediterranean cyclones' centers from 1979 to 2020.
The dataset consisted of $368'184$ elements (wind intensity snapshots), 
whose features were the weights of LDA topics.
With $20$ to $30$ motifs, this represents a reduction in size of $\sim\!\!10^3$ times with respect to the original ERA5 dataset. 

Different machine learning techniques were tested, belonging to two main families. 
In all cases, the dataset was randomly split into a training and a test set with a ratio of $4\!:\!1$. Let us emphasize that this completely disregarded the temporal structure of the single cyclones, as each snapshot was considered on its own.
For all the referred algorithms we used their implementation in the \texttt{sklearn} Python library.

\subsection{Classification approach}

We considered a grid dividing the target area in a fixed number ($30\times 15$) of equally spaced elements (plus one representing the no-cyclone case) and tried to associate a single wind mask with the center of a cyclone in one of these zones. 
We tested quadratic classifiers, $k$-nearest neighbor classifiers and Support Vector Classification.
The latter yielded the best results using a polynomial kernel of degree $11$ (tested with degrees 2-15, non-polynomial kernels lead to worse results). This approach led to a level of accuracy of $82\%$, with a substantial part of the errors being due to cyclones predicted in a zone adjacent to the right one.

This naturally led to the second approach, that is \textbf{regression techniques} based on the distance between actual cyclones and predicted ones. We tested a Multi-Layer Perceptron Regressor (MLP) with multiple choices of the parameters: number of LDA topics, width and depth of the neural network.
For snapshots of the dataset that did not include any cyclone (most of them), a fictitious cyclone was added at coordinates well outside of the Mediterranean region under consideration, and we interpreted as ``no cyclone present'' any output of the MLP on the test set that consisted in coordinates outside of that region.
The best results were given by the LDA with $25$ topics, which we report in \cref{tab:r2}, and the optimal choice turned out to be a network with $4$ layers of $800$ neurons each.

\begin{table}
\caption{\label{tab:r2}$R^2$ scores obtained using a Multi-Layer Perceptron. Results in the same column share the same number of layers, while rows are ordered by layer width.}
\begin{ruledtabular}
\begin{tabular}{c|cccc}
\multicolumn{5}{c}{$R^2$ scores of MLP}\\
\hline
\rule{0pt}{2.5ex}&4&5&6&7\\
\hline
\rule{0pt}{2.5ex} 500     &0.889831   &0.900812   &0.901976   &0.891367 \\
600     &0.899367   &0.896501   &0.901746   &0.895966 \\
700     &0.902432   &0.915464   &0.912656   &0.903862 \\
800     &\textbf{0.915524}   &0.916272   &0.909754   &0.898197 \\
900     &0.913419   &0.918152   &0.917083   &0.913861 \\
1000    &0.919801   &0.901976   &0.924494   &0.914722 \\
\end{tabular}
\end{ruledtabular}
\end{table}


\section{Conclusions}

We have obtained evidence of the efficacy of a supervised machine learning approach for the detection of Mediterranean cyclones based on the output of a Latent Dirichlet Algorithm applied to wind data. The Bayesian LDA algorithm allowed a drastic dimensional reduction.
Moreover, the procedure shows how cyclones can in fact be operationally characterized with a relatively small number of features, which might be interesting also in the scope of a theoretical analysis of the phenomenon.

Detection of Mediterranean cyclones may be approached with other machine learning algorithms, for instance with Convolutional Neural Networks often used in image recognition and processing. One major obstacle is that data for training the model consist in the tracks of the cyclones' centers, and there is no reliable procedure for determining the shape or diameter of a cyclone, for the same reasons that make it difficult to provide a rigorous characterization of such phenomena. Our Bayesian (LDA) approach overcomes this difficulty identifying the typical structures concurring in a cyclone, albeit in a rather implicit manner. 

A natural way to further our investigation should involve the inclusion of the temporal structure of Mediterranean cyclones in the LDA analysis, with the aim of detection of \emph{precursors} of cyclones, especially of extreme ones. On the other hand, our approach easily lends itself to the analysis of many other non-autonomous geophysical systems: as it was showcased in the detection of cyclones, it allows to robustly identify trends and structures typically appearing in a system responding to external perturbations.

\begin{acknowledgments}
Authors F. G. and L. R. were supported by the project \emph{Mathematical methods for climate science}, funded by he Ministry of University and Research (MUR) as part of the PON 2014-2020 ``Research and Innovation'' resources - Green Action - DM MUR 1061/2022. F. G. and L. R. completed this work during their collaboration with Miningful srls.

Computations have been performed on the computing cluster Toeplitz of the Department of Mathematics at University of Pisa.
The authors wish to thank Andrea Agazzi, Nevio Dubbini and Davide Faranda for many insightful suggestions.
\end{acknowledgments}



\section*{Data Availability Statement}

The data that support the findings of this study are available from the corresponding author upon reasonable request.

\section*{References}

\nocite{*}
\bibliography{biblio}

\end{document}